\documentclass[%
 reprint,
 amsmath,amssymb,
 aps,
]{revtex4-1}

\usepackage{graphicx}% Include figure files
\usepackage{dcolumn}% Align table columns on decimal point
\usepackage{bm}% bold math
\allowdisplaybreaks

\begin{document}

\title{Intensities of $\gamma$-ray emissions following $^{111}Sn$ 
decay determined via photonuclear reaction yield measurements}

\author {A. Chekhovska}
\email{chekhovska@kipt.kharkov.ua}
\affiliation{V. N. Karazin Kharkiv National University, 4 Svobody Sq., 
Kharkiv, 61022, Ukraine}
\affiliation{Institute of High-Energy and Nuclear Physics of NSC 
"Kharkiv Institute of Physics and Technology", 1 Akademichna St., Kharkiv,
61108, Ukraine}
\author {Ye. Skakun}
\author {I. Semisalov}
\author {S. Karpus}
\author {V. Kasilov}
\affiliation{Institute of High-Energy and Nuclear Physics of NSC 
"Kharkiv Institute of Physics and Technology", 1 Akademichna St., Kharkiv,
61108, Ukraine}

\date{\today}

\begin{abstract}
Intensities of ten strongest $\gamma$-ray transitions following $^{111}Sn$ 
$(T_{1/2}$=35.3 $min$) decay have been determined via comparison of
two sets of the experimental photonucleon reaction yields
driven  using traditional activation equation and activation equation for 
genetically coupled radioactive nuclei. The found absolute intensities of the 
$\gamma$-ray transitions in question turned up to be noticeably different 
from the currently recommended values.
\end{abstract}

\maketitle

\section{Introduction}
\label{Introduction}

\begin{figure} [h]
    \includegraphics[width = 80 mm]{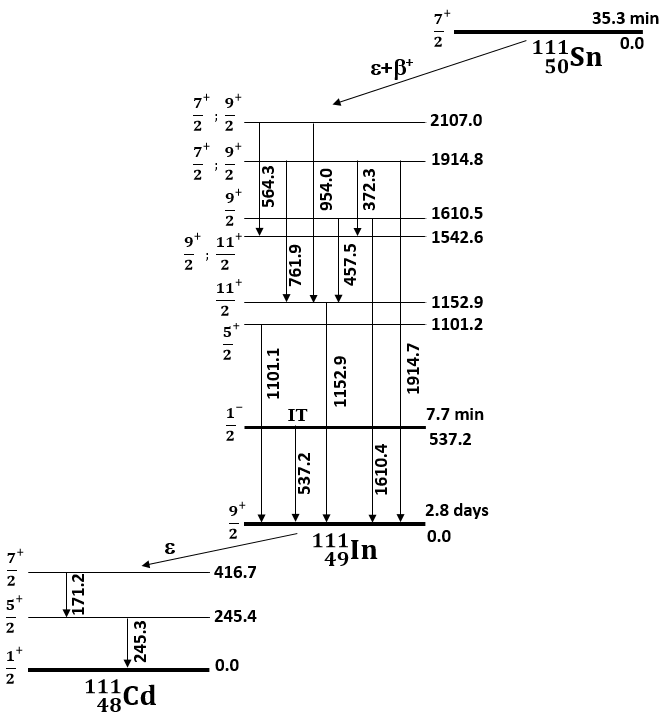}
	\caption{Simplified scheme of the $^{111}Sn \, \to 
	\, ^{111m,g}In \, \to \, ^{111}Cd$ radioactive chain.}
	\label{FIG:1}
\end{figure}

Nucleus decay data are important for both nuclear 
spectroscopy theories and experimental techniques 
for defining nuclear reaction cross sections or yields 
by means of residual activity measurements. 
The tin-111 ($^{111}Sn$) nucleus decaying via 
$(\varepsilon+\beta^+)$-process with the half-life of 
35.3 $min$ populates a 
large array of excited levels of the indium-111 ($^{111}In$) 
daughter nuclide among which 
there is an isomeric state with the excitation energy 
537.2 $keV$ and the
half-life $T_{1/2}^{m}$=7.7 $min$ (Fig.~\ref{FIG:1}). 

\noindent
The $^{111}In$ ground state $(T_{1/2}^{g}$=2.80\ 
$days)$ decays to the stable $^{111}Cd$ 
one following the strong $\gamma$-ray
transitions of 171.2 $keV$ and 245.3 $keV$ energies. 
The latest evaluated decay data for  
A=111 nuclear mass were recommended in the work
\cite{Blachot:2009ujo} and 
included to NuDat 2.8 base \cite{nudat2}. Meanwhile the
intensity values of $\gamma$-ray transitions between $^{111}In$ 
excited levels following 
the $^{111}Sn$ decay are based on relatively old 
experimental measurements,
mostly performed in the 1970-1980s (see references 
in \cite{Blachot:2009ujo}), 
with the use of detectors of relatively low efficiency 
and poor resolution 
compared with current $\gamma$-ray spectrometry techniques. 

A large quantity of experimental measurements of 
activation cross sections 
and yields of different nuclear reactions induced 
by various incident particles, 
which lead to the formation of the $^{111}Sn$ nuclide, 
has been carried out 
for various basic and applied purposes \cite{exfor}.
Correct values of $\gamma$-ray emissions following 
residual nuclei 
are needed for right determination of
nuclear reaction cross sections or yields using 
the $\gamma$-ray spectrometry
activation technique. We faced this problem while
determining the bremsstrahlung
activation yields of the near-threshold photonuclear 
reactions with the $^{112}Sn$
nuclide as a target which are partly of interest as 
input data for studying the 
$\gamma$-scenario of stellar nucleosynthesis 
of the so-called $p$-nuclei
\cite{Rauscher_2013,Rauscher_2018}.

\section{Experimental procedure and analysis}
\subsection {General notes}

\begin{figure*} 
    \includegraphics[width = 175 mm]{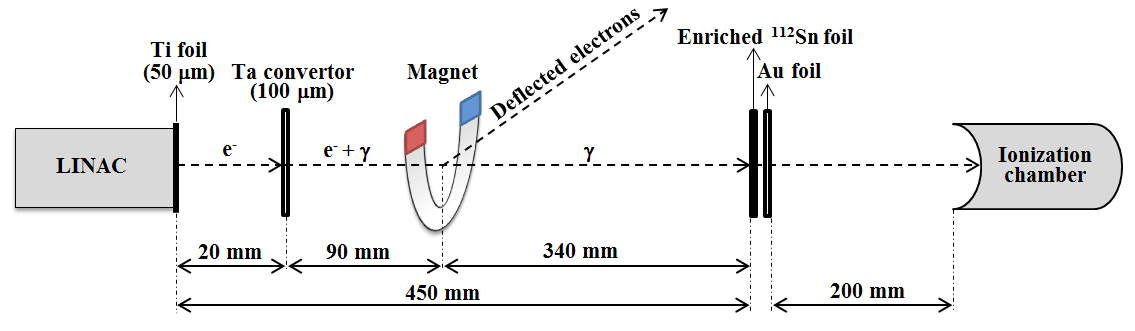}
	\caption{Sketch of the target irradiation at LINAC.}
	\label{FIG:2}
\end{figure*}

The $^{111}Sn$ radioisotope was produced by the
$^{112}Sn(\gamma,$n)$^{111}Sn$
photonuclear reaction at the 30 $MeV$ electron 
\textbf{LIN}ear \textbf{AC}celerator (LINAC) located at National 
Science Center "Kharkiv Institute of Physics and Technology" (NSC KIPT). 
Fig.~\ref{FIG:2} illustrates the scheme of our experimental
setup for irradiation of targets. 

\begin{figure} [h]
    \includegraphics[width = 85 mm]{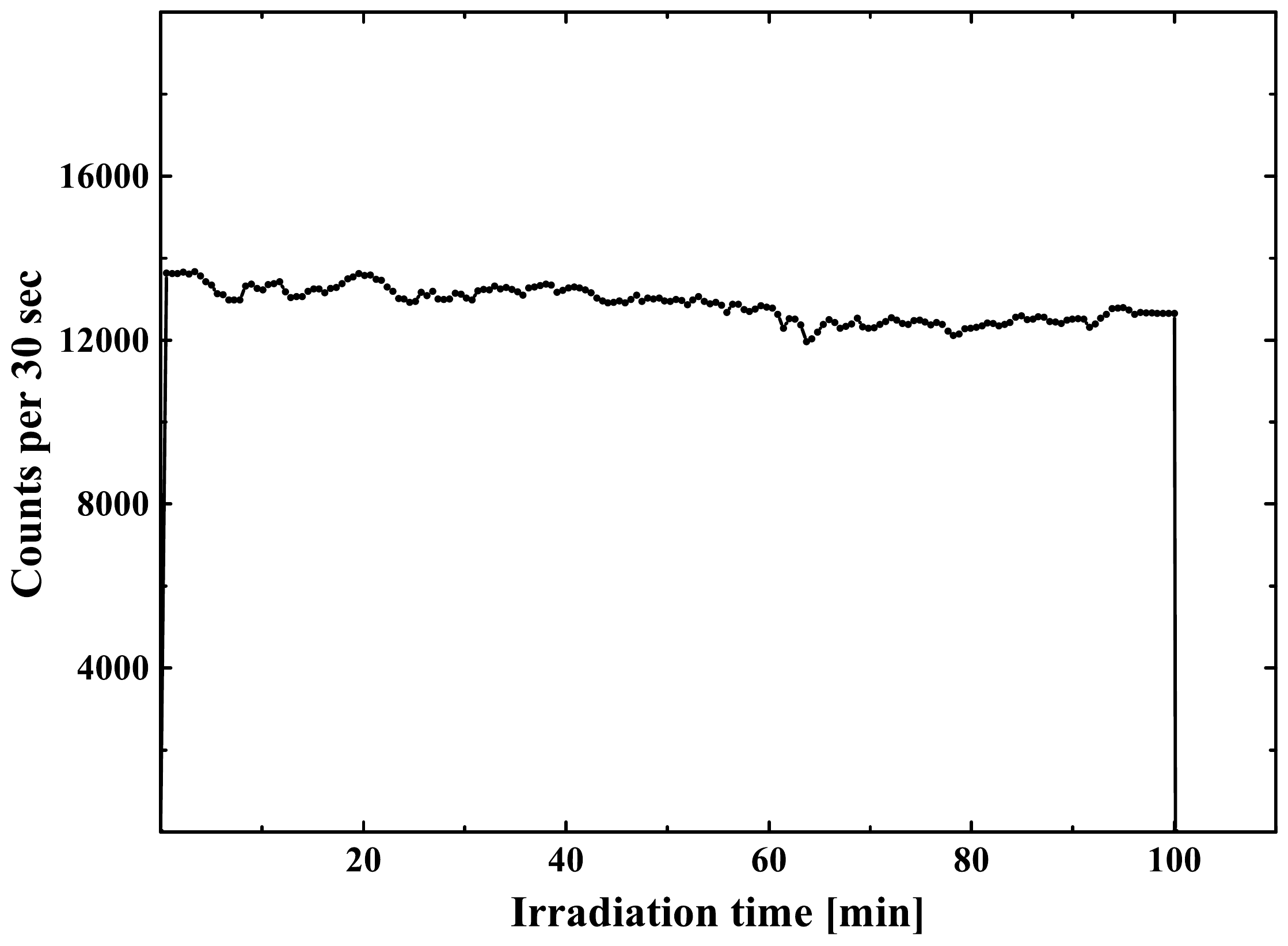}
	\caption{Typical example of the photon flux intensity 
	in the target during 100 $min$ 
	irradiation monitored with the ionization chamber.}
	\label{FIG:3}
\end{figure}

The electron beam of about 10 $\mu A$ average 
operating current having 15 $MeV$ and less energies was 
deflected by the angle of 
35$^{\circ}$ with a sector magnet (not shown in
Fig.~\ref{FIG:2}) creating 2\% 
energy half-width of the beam. Having passed through 
the 50 $\mu m$ titanium window the beam impacted 
at the 100 $\mu m$ tantalum foil to be converted into a
bremsstrahlung photon flux 
irradiating the investigated targets placed along 
the electron beam axis while
the remained electrons were deflected by a permanent magnet.
Four self-supporting tin metallic foils having the 
square shape with side 
18 $mm$ and total weight of 77 $mg$ enriched with the
$^{112}Sn$ isotope to 80\%  
were used as a unified target. At every irradiation 
the gold foil of 20 $mm$ 
in diameter with the weight of 120 $mg$ was placed with the 
studied tin target in the 
close geometry in order to use the
$^{197}Au(\gamma,$n$)^{196}Au$ reaction 
as a standard one to determine the bremsstrahlung flux. 
The cross sections 
of the last reaction had been earlier measured and 
evaluated by several
experimental teams
\cite{article1,troshchiev2010new-data1108606,
Nair_2008,article2} 
in the giant resonance region. Their results are well
consistent with each other and the $^{196}Au$ 
residual radioactive decay has 
quite suitable properties \cite{nudat2} for its 
activity measurement. 
Several exposures of the combined target 
(the target sandwich) were 
carried out over the range 
of the bremsstrahlung endpoint energies between the 
threshold of the 
$^{112}Sn(\gamma$,n)$^{111}Sn$ reaction (10.79 $MeV$) 
and 15 $MeV$ to obtain the
energy dependence of the photoactivation yield. 
The ionization chamber placed along 
the beam axis and screened from the background 
radiation with a lead shield was monitoring the
bremsstrahlung photon flux by regular recording the
X-ray dose rate. 
Fig.~\ref{FIG:3}  shows the example of photon flux 
intensity as a function of time during irradiation. 
Necessary corrections were taken into account for 
the reaction yield calculations in the cases
of essential photon flux fluctuations. 

Irradiation lasted 60 and 100 $min$ (to determine the
$^{112}Sn(\gamma,$n)$^{111}Sn$ reaction yields) and 
20 $min$ (to determine the $^{112}Sn(\gamma,$p)$^{111m}In$
reaction yields). After each irradiation the targets were
instantly delivered 
to a low-background room far from the accelerator in order
to begin the measurements of energy spectra of $\gamma$-rays
following the 
radioactive decay of $^{111}Sn$ and its daughter 
$^{111}In$ using a coaxial 
Canberra High Purity Germanium (HPGe) detector with the
relative efficiency of 30\% 
in comparison to the efficiency of (3 in.×3 in.)
$NaI(Tl)$-detector and 1.8 $keV$ 
resolution for the 1332 $keV$ $\gamma$-line of the 
$^{60}Co$ isotope source. 
To reduce ambient radioactivity the detector was contained 
in a lead shield, with the walls of 12 $cm$ in width and 
degraders of 3 $mm$ $Cd$ and 
5 $mm$ $Cu$ line inside the shield to reduce the 
interference of the $Pb$ fluorescence X-rays. 
The $\gamma$-ray spectra of 
$^{196}Au$ $(T_{1/2}$ = 6.16 $days$ \cite{nudat2}) 
residual nucleus of the 
standard reaction were measured secondarily. 
The irradiated targets were 
mounted along the vertical axis of the spectrometer 
at several sample-to-crystal 
distances between 5 and 10 $cm$. The final measurements 
were carried out at 
sample-to-crystal distances providing 2\% or less 
dead time and a 
negligible summing effect.

\begin{figure} [h]
    \includegraphics[width = 85 mm]{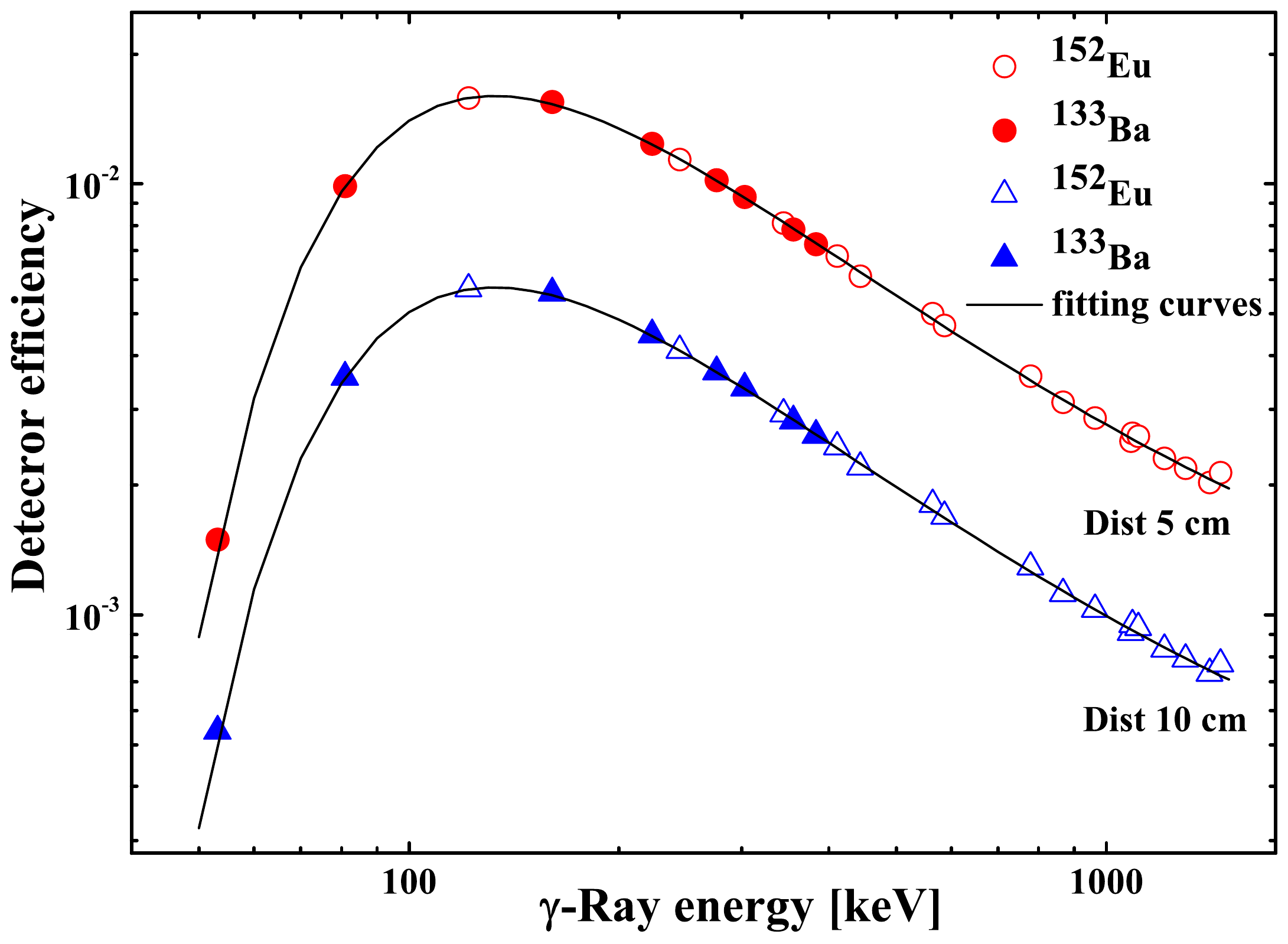}
	\caption{Full-energy peak detection efficiency 
	curves of HPGe
     $\gamma$-ray spectrometer for sample-to-endcap 
     distances 5 and 10 $cm$.}
	\label{FIG:4}
\end{figure}

The measurements of the detector full-energy-peak 
efficiency were performed in the (50-1500) $keV$ 
$\gamma$-ray energy region using 
$^{133}Ba$ and $^{152}Eu$ calibrated point sources.
Fig.~\ref{FIG:4} 
shows the energy dependences of the detector efficiency 
for two 
distances between the source and the crystal endcap.

\subsection {Activity measurements}

\begin{figure} [h]
       \includegraphics[width = 85 mm]{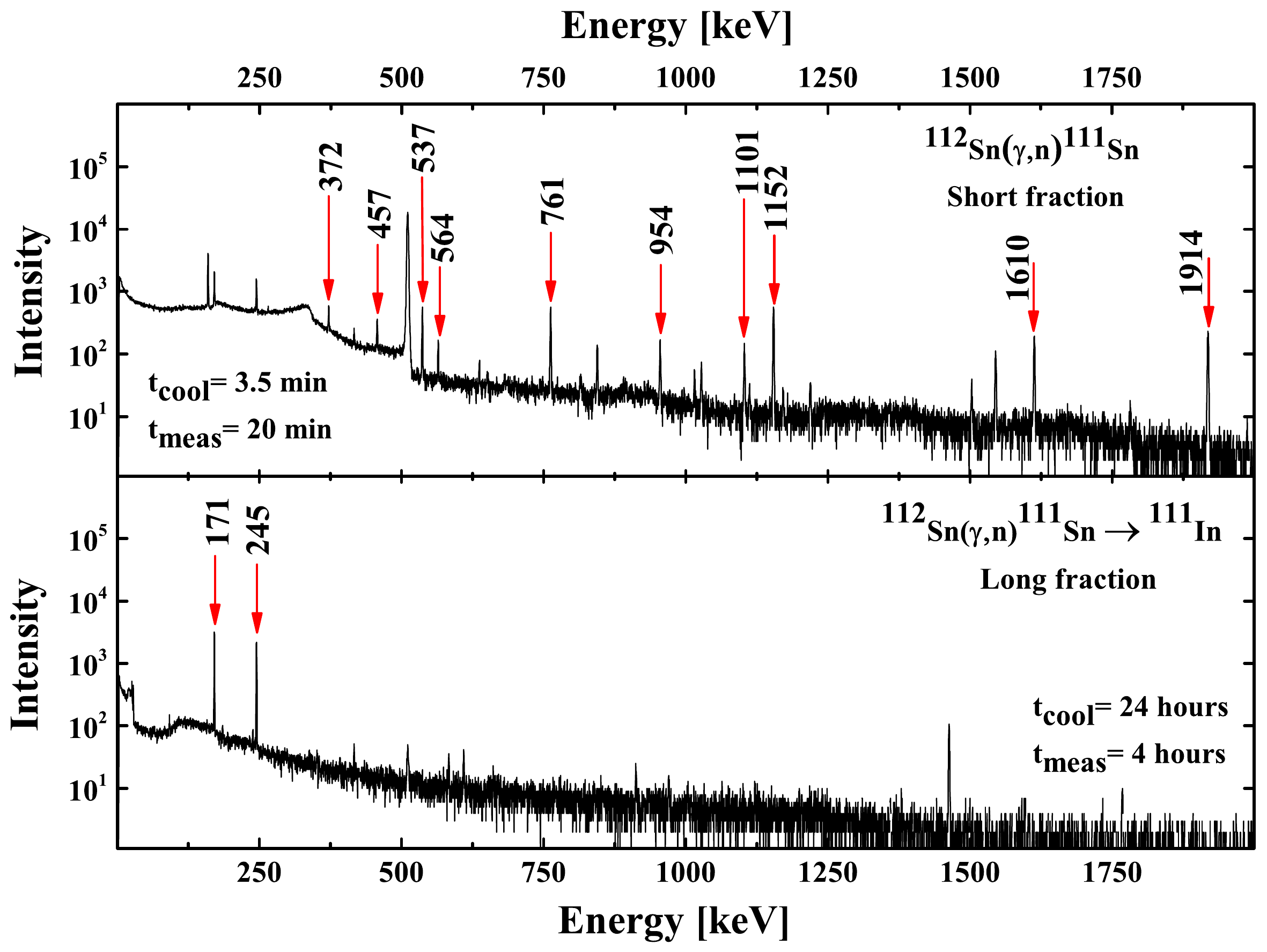}
       \caption{Short (upper panel) and long (lower panel)
       fractions of the typical 
         $\gamma$-ray spectrum measured after irradiation of
         the $^{112}Sn$ target 
         with 15 $MeV$ bremsstrahlung.}
	   \label{FIG:5}
\end{figure}

Two typical $\gamma$-ray spectra of the $^{112}Sn$ 
target irradiated with 
15 $MeV$ bremsstrahlung are shown in Fig.~\ref{FIG:5}. 
The short-live fraction 
of the induced radioactivity is given in the upper panel, 
the long-live one 
in the lower panel. The arrows of the upper panel indicate 
10 strongest gamma-ray
transitions in the $^{111}In$ nucleus following the 
$^{111}Sn$ decay. 

\noindent
The energy of each transition is indicated in
kiloelectron-volt units above
the arrow. A number of other weaker $\gamma$-ray 
emissions following the 
$^{111}Sn$ and other side radioisotopes, in particular 
$^{123m}Sn$ $(T_{1/2}$=40.06 $min$, $E_{\gamma}$=160.3 $keV$),
can be identified in this $\gamma$-ray spectrum as well.
The $\gamma$-ray spectrum measured one day after irradiation
(the lower panel of
Fig.~\ref{FIG:5}), except the weak background, contains 
only 2 strong peaks (171.2 $keV$ 
and 245.3 $keV$, indicated with arrows), which 
correspond to the $\gamma$-rays 
following the decay of the $^{111g}In$ nucleus 
$(T_{1/2}^{g}$ = 2.80 $days$) 
being the daughter of the $^{111}Sn$ nucleus (see
Fig.~\ref{FIG:1}) and
on the other side can be additionally produced via the 
$^{112}Sn(\gamma,$p)$^{111m+g}In$ reaction (the 7.55 $MeV$
threshold) 
in appliance with the scheme:

\begin{figure} [h]
		\includegraphics[width = 85 mm]{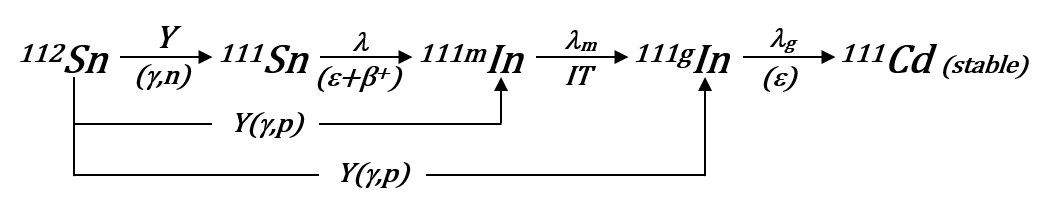}
	    \label{Scheme}
\end{figure}

Energies and intensities of the mentioned $\gamma$-ray 
transitions of 
the $^{111}Sn \, \to \,^{111}In \, \to \,^{111}Cd$ 
radioactive chain borrowed 
from NuDat 2.8 base \cite{nudat2} are presented in
Table~\ref{tbl1}.
 
{\renewcommand{\arraystretch}{1.4}% for the vertical padding
\begin{table} [h]
    \caption{Energies and intensities of $\gamma$-ray
    transitions following 
    $^{111}Sn$,$^{111m}In$, and $^{111g}In$ decays
      \cite{nudat2}}
    \label{tbl1}    
    \centering
    \begin{tabular}{|c|c|}
        \hline
        \textbf{$E_{\gamma}$ [$keV$]} &
        \multicolumn{1}{c|}{$I_{\gamma}$ [\%]} \\ 
        \hline
        \multicolumn{2}{|c|}{$^{111}Sn \, \to \, ^{111}In$}\\
        \cline{1-2}
        372.3 & 0.42 (7)  \\ 
        457.5 & 0.38 (6)  \\ 
        537.2 & 0.25 (4)  \\ 
        564.3 & 0.30 (5)  \\ 
        761.9 & 1.48 (23)  \\ 
        954.0 & 0.51 (8)  \\ 
        1101.1 & 0.64 (11) \\ 
        1152.9 & 2.7   \\ 
        1610.4 & 1.31 (20)  \\ 
        1914.7 & 2.0 (3)  \\ 
        \hline
        \multicolumn{2}{|c|}{$^{111m}In \, \to \,
        ^{111g}In$}\\ \cline{1-2}
        537.2 & 87.2 (5)   \\ 
        \hline
        \multicolumn{2}{|c|}{$^{111g}In \, \to \, ^{111}Cd$}\\
        \cline{1-2}
        171.2 & 90.7 (9)  \\
        245.3 & 94.1 (10)  \\
        \hline
    \end{tabular}     
\end{table}}

\section {Experimental data analysis}

\begin{figure} [h]
    \includegraphics[width = 85 mm]{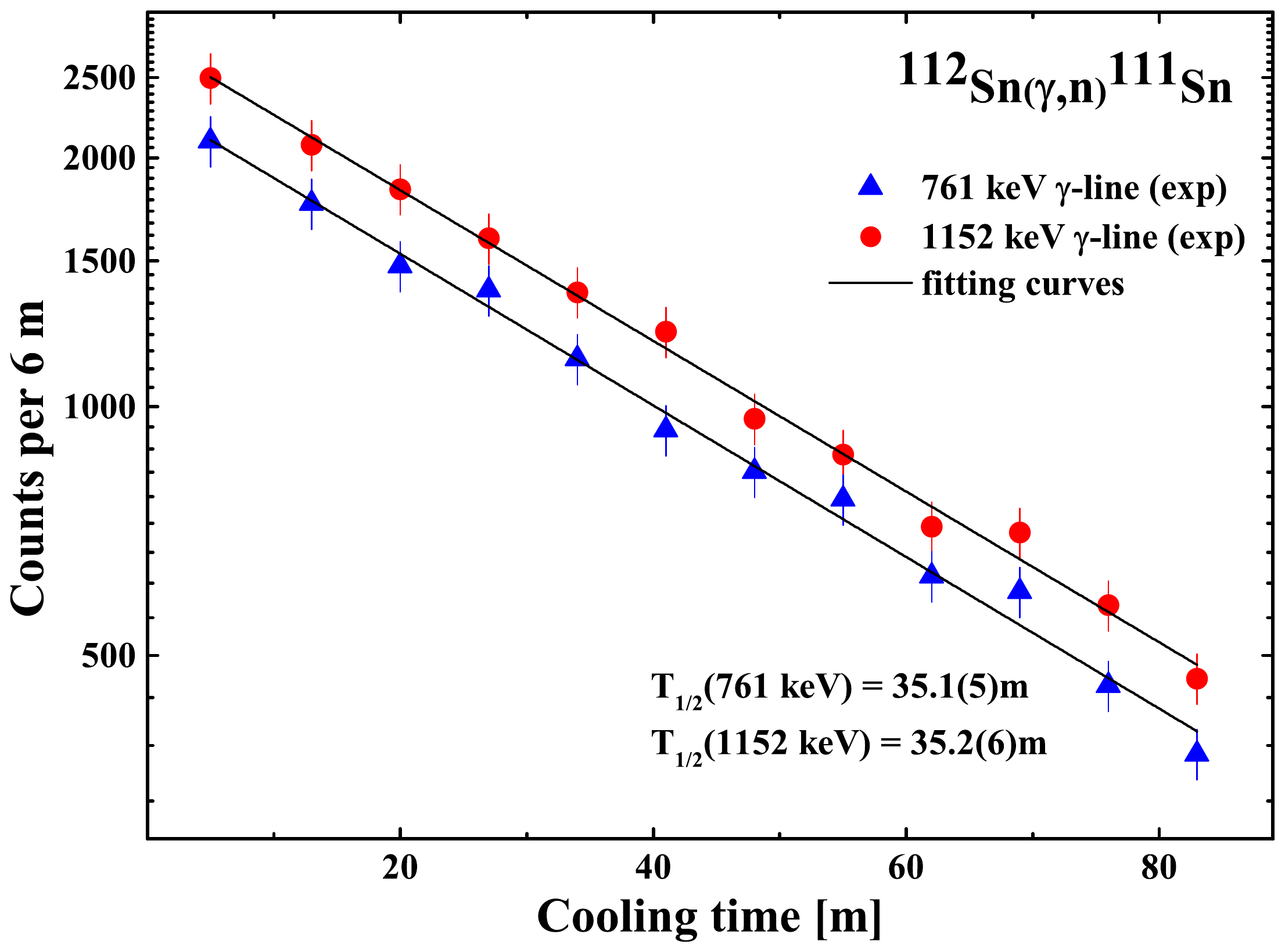}
	\caption{Decay curves of the $^{111}Sn$ radioactive
	nucleus constructed from 761 $keV$ and 1152 $keV$
	$\gamma$-line intensities.}
	\label{FIG:6}
\end{figure}

Radioactive decay curves derived analyzing the two 
most intense 
$\gamma$-ray transitions (761.9 and 1152.9 $keV$) of the
$^{111}In$ daughter 
nucleus are depicted in Fig.~\ref{FIG:6}. The half-life 
values 
(indicated in the plot) of the $^{111}Sn$ radionuclide
determined from the time
dependencies of the intensities of these two 
$\gamma$-lines are in good 
agreement with NuDat 2.8 base value 35.3 (6) $min$
\cite{nudat2}. 
Remaining gamma lines following the $^{111}Sn$ decay 
obey the same 
consistent pattern of exponential decay.

The bremsstrahlung activation yield $Y$ of the
$^{112}Sn(\gamma,$n$)^{111}Sn$ 
reaction can be determined via solution of the traditional 
activation equation~(\ref{equ1}):
\begin{equation}
\begin{aligned}
\frac{S_{\gamma}}{\varepsilon \cdot Br \cdot n \cdot \phi(E_{j})}=
\frac {Y}{\lambda} \cdot \bigl(1 - e^{-\lambda t_{1}}\bigr) 
\cdot e^{-\lambda t_{2}} \cdot \bigl(1 - e^{-\lambda
t_{3}}\bigr)
\end{aligned}
\label{equ1}
\end{equation}
in which $S_{\gamma}$ is the experimental area of any
$\gamma$-ray peak 
of the $^{111}Sn$ decay, $\varepsilon$ is the full-energy peak
detection efficiency,
$Br$=$I_{\gamma}/100$ is the branching coefficient of the same
$\gamma$-ray transition, 
$n$ is the number of nuclei in the target being irradiated,
$\phi(E_{j})$ is the $E_{j}$ end-point energy bremsstrahlung
fluence covering the target, $\lambda$ is
the radioactive decay constant, 
$t_{1}, t_{2},$ and $t_{3}$ are irradiation, cooling and
measurement times of the target activity respectively.

The fluence $\phi(E_{j})$ of the bremsstrahlung, penetrating 
the sandwiched targets of the studied $^{112}Sn$ foil and the
$^{197}Au$ standard one, can be determined from the 
experimentally measured $^{197}Au(\gamma$,n)$^{196}Au$ reaction
photoactivation yield $Y_{Au}(E_{j})$, the angle-integrated
bremsstrahlung spectrum $W(E,E_{j})$ \cite{PhysRev.83.252} 
and the cross section
energy dependence $\sigma_{Au}(E)$ of the standard reaction
\cite{article1} by the equation~(\ref{equ2}):
\begin{equation}
  \begin{aligned}
\phi(E_{j})= \frac{Y_{Au}(E_{j})}{\int_{E_{th}}^{E_{j}}W
(E,E_{j})\cdot\sigma_{Au}(E)dE}.
   \end{aligned}
   \label{equ2}
\end{equation}

De-excitation of the $^{111}In$ states (including the
$^{111m}In$ isomer) 
populated by the $^{111}Sn$ nucleus decay leads to the 
$^{111}In$ ground state. 
The experimental areas $S_{\gamma}$ of 171.2 $keV$ 
and 245.3 $keV$ 
$\gamma$-ray peaks of the $^{111g}In$ decay with the 
cooling time of $t_{2}$ being much longer than 7.7 
$min$ (the $^{111m}In$ isomer half-life) 
obey the equation~(\ref{equ3}) \cite{Friedlander1981} 
for genetically-coupled radioactive nuclides:
\begin{equation}
  \begin{aligned}
 \frac{S_{\gamma}} {\varepsilon \cdot Br \cdot n \cdot
 \phi(E_{j})}  = \\
 Y_{p} \cdot \frac{\lambda_{p} \cdot \lambda_{d}}{\lambda_{d} - \lambda_{p}} 
 \cdot \bigl[\frac{1-e^{-\lambda_p t_{1}}}
 {\lambda^2_p} \cdot e^{-\lambda_p t_{2}} \cdot \bigl
 (1 - e^{-\lambda_p t_{3}}\bigr) \\
 - \frac{1-e^{-\lambda_d t_{1}}}{\lambda^2_d} \cdot
 e^{-\lambda_d t_{2}} \bigl(1 - e^
 {-\lambda_d t_{3}}\bigr) \bigr] \\
 + Y_{d} \frac {1-e^{-\lambda_d t_{1}}}{\lambda_d} \cdot
 e^{-\lambda_d t_{2}} 
 \cdot \bigl(1 - e^{-\lambda_d t_{3}}\bigr),
   \end{aligned}
   \label{equ3}
\end{equation}
 where in our case $Y_{p}$ ($Y$ in the traditional activation
 equation~(\ref{equ1})) 
 and $Y_{d}$ are the yields and
 $\lambda_{p}$ and $\lambda_{d}$ are the decay constants 
 of the \textbf{p}arent $^{111}Sn$
 and \textbf{d}aughter $^{111}In$ nuclei respectively.

\begin{figure} [h]
    \includegraphics[width = 85 mm]{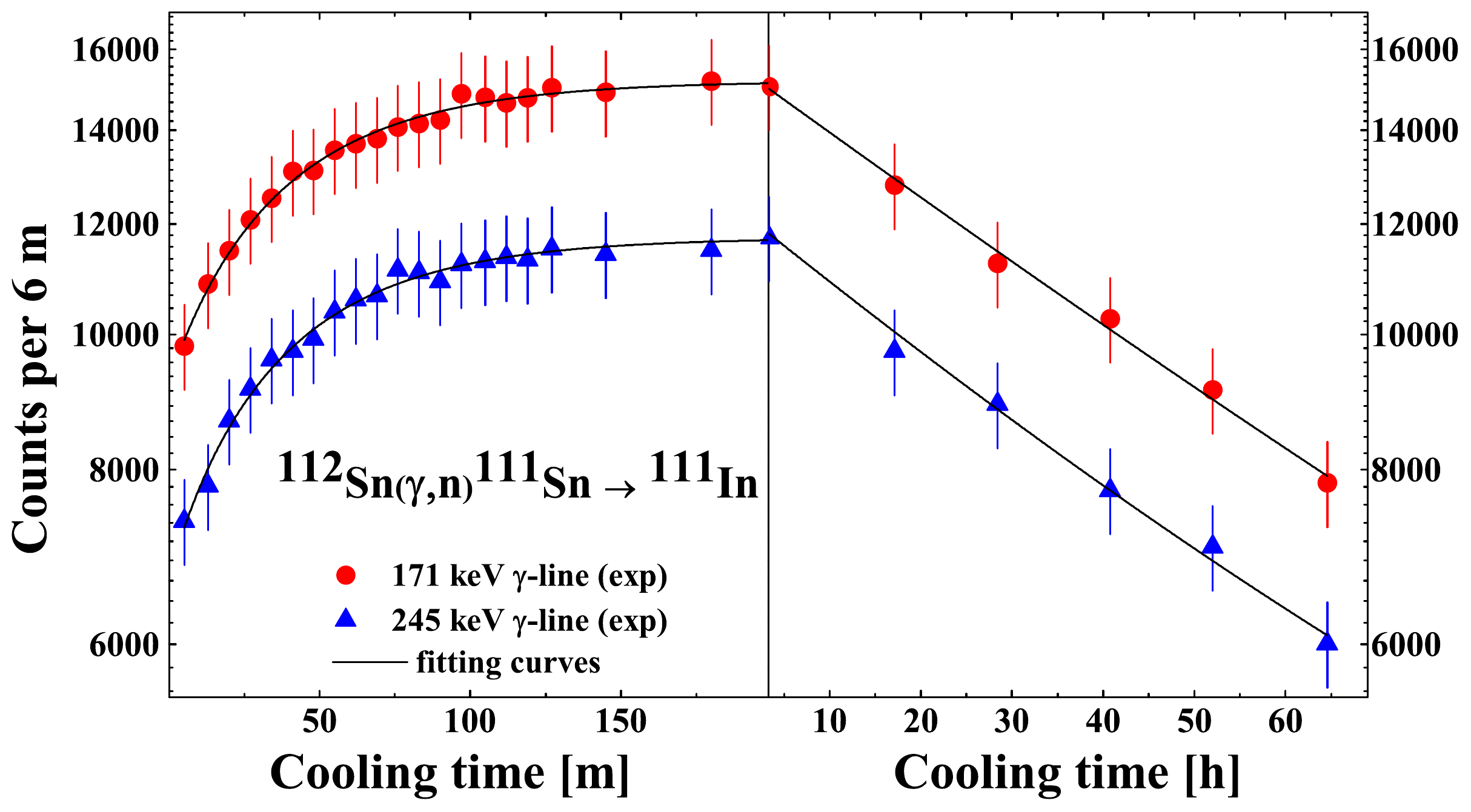}
	\caption{Accumulation and decay curves of the 
	$^{111}In$ isotope nuclide produced by the 
	$^{112}Sn(\gamma$,n)$^{111}Sn \to \, ^{111}In$ process.}
	\label{FIG:7}
\end{figure}

\begin{figure} [h] 
    \includegraphics[width = 85 mm]{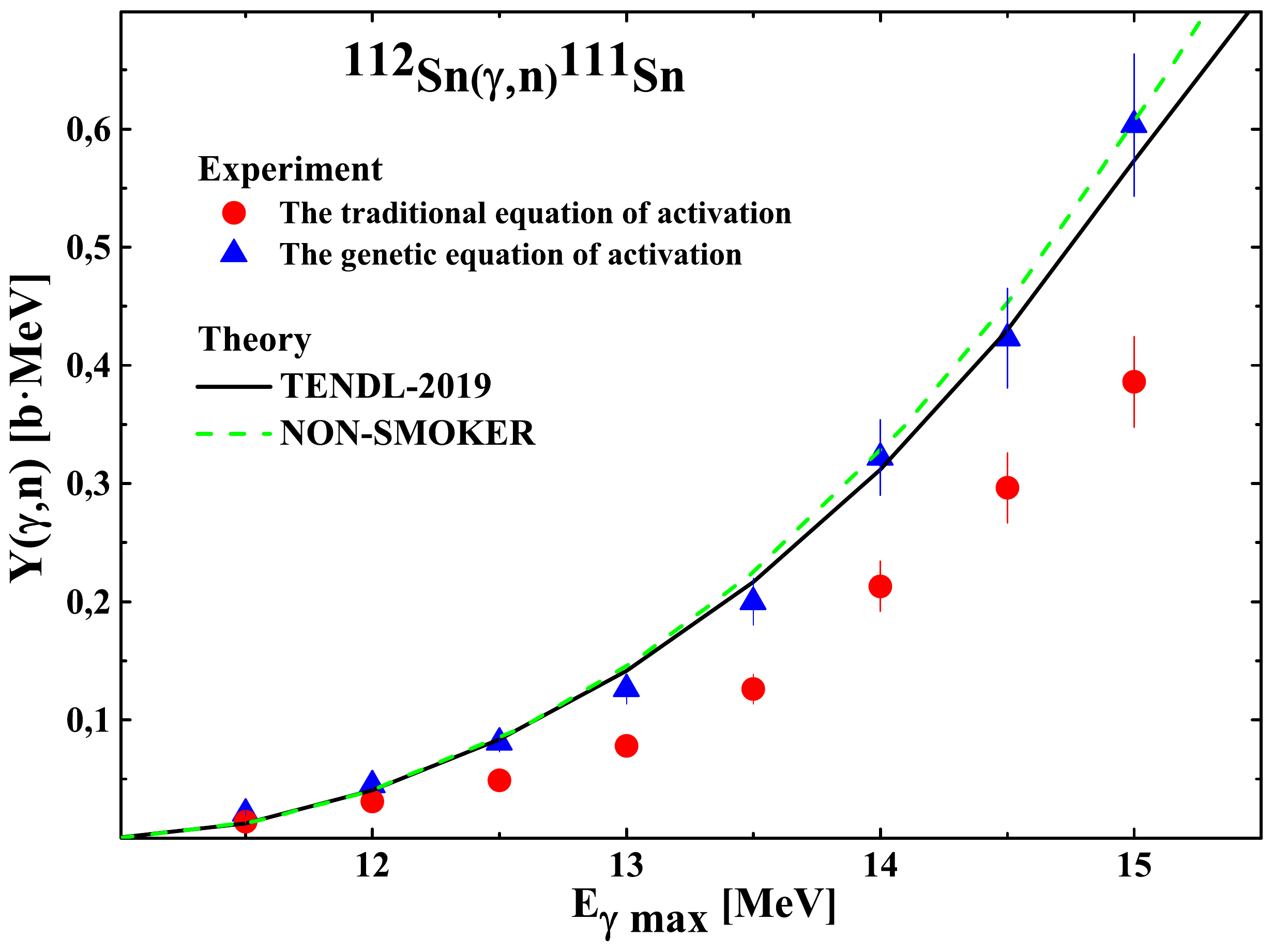}
	\caption{The $^{112}Sn(\gamma$,n)$^{111}Sn$ reaction
	yields determined 
	using the traditional activation equation (circles) 
	and equation for 
	genetically coupled activities (triangles).}
	\label{FIG:8}
\end{figure}

The curves of the $^{111}In$ nucleus accumulation and 
decay, plotted according to 171.2 $keV$ and 245.3 $keV$
experimental $\gamma$-line intensities,
measured after the end of irradiation of the tin target, 
are shown in Fig.~\ref{FIG:7}. The forms of these time
dependencies are due 
to the differences of the half-lives of the parent and
daughter members of the radioactive chain and the 
values of the activation yields ($Y_p$ and $Y_d$ in the equation~(\ref{equ3})) 
of the $^{112}Sn(\gamma,$n)$^{111}Sn$ 
and $^{112}Sn(\gamma$,p)$^{111}In$ reactions respectively. 
The increasing parts of the $^{111}In$ activity curves at 
the left part of Fig.~\ref{FIG:7} are explained by the 
feeding of the longer-living nucleus by the 
shorter-living one decay. 
Fitting the equation~(\ref{equ3}) for genetically 
coupled activities by least squares method to the experimental
points we were able to determine the independent
values of the both activation yields and obtained an 
unexpected result: the values of $Y_p$ (i. e. the yields 
of the $^{112}Sn(\gamma,$n)$^{111}Sn$ reaction) turned 
out to be noticeably larger than those determined using 
the traditional activation equation~(\ref{equ1}) and 
the NuDat 2.8 base data \cite{nudat2} for the $^{111}Sn$ 
decay $\gamma$-ray intensity values. Both data sets 
for different 
bremsstrahlung energies are shown in Fig.~\ref{FIG:8} 
where the circles represent the experimental weighted 
average photonuclear
$^{112}Sn(\gamma$,n)$^{111}Sn$ reaction yields 
calculated applying the traditional activation 
equation~(\ref{equ1}) and the current \cite{nudat2}
$\gamma$-ray emission values of 10 strongest 
$\gamma$-ray transitions of the $^{111}Sn \, \to \, ^{111}In$
decay. The values of the triangle form points were 
obtained applying 
equation~(\ref{equ3}) for genetically coupled activities 
and the database \cite{nudat2} emission values of the 
171.2 $keV$ and 245.3 $keV$ $\gamma$-rays of the 
$^{111}In \, \to \, ^{111}Cd$ decay.

The decay characteristics (including $\gamma$-ray emission
intensities) of the long-lived $^{111}In$ ground state nucleus
have been investigated quite well by now and the only 
reason for this observation may be hidden in large uncertainties 
of the $\gamma$-ray emission 
values of the radiation transitions, following the 
$^{111}Sn$ nucleus decay, evaluated earlier \cite{nudat2} 
from the energy and intensity balances of the branches 
populating and de-exciting the $^{111}In$ levels. 
The $\gamma$-ray intensities of the $^{111}Sn$ radionuclide 
decay presented in work \cite{Blachot:2009ujo} were normalized
based on: 
1) total probabilities of radiation transitions (i.e.
$\gamma$-rays + internal conversion electrons), 
2) electron capture probability to the $^{111}In$ ground state, and 
3) $\beta^{+}$-decay probability to the $^{111}In$ ground state.
Each of these probabilities was measured by a separate method and
has its own error.

However the favorable features of the 
$^{111}Sn \, \to \, ^{111}In \, \to \, ^{111}Cd$ radionuclide
chain gave us possibility to
determine the reliable experimental independent activation yields
of the $^{112}Sn(\gamma$,n)$^{111}Sn$ photonuclear reaction using
activation equation~(\ref{equ3}) for genetically coupled
radioactive nuclides and 171.2 $keV$ and 245.3 $keV$
$\gamma$-ray line experimental areas with only gamma-spectrometry
technique uncertainties (less than 2\% of counting statistics, 
1\% of branching coefficients \cite{nudat2} and 5\% of detector
efficiency). Comparing these activation yields with those
determined via solution of traditional activation
equation~(\ref{equ1}) we were able to derive updated values 
of the $^{111}Sn$ nuclei emission intensities. 
They are presented in the right column of Table~\ref{tbl2}.

 {\renewcommand{\arraystretch}{1.4}% for the vertical padding
 \begin{table} [h]
    \caption{New intensities of $^{111}In$ $\gamma$-ray
    transitions 
    following $(\varepsilon+\beta^{+})$-decay of the
    $^{111}Sn$ nucleus.}
    \label{tbl2}    
    \centering
    \begin{tabular}{|c|c|c|}
        \hline 
        $E_\gamma$ & \multicolumn{2}{c|}{$I_\gamma$ [\%]}\\
        \cline{2-3}
        [$keV$]     & NuDat\cite{nudat2} & New data \\
        \hline
        372.31 & 0.42 $\pm$ 0.07 & 0.26 $\pm$ 0.02 \\
        457.56 & 0.38 $\pm$ 0.06 & 0.23 $\pm$ 0.02 \\
        537.20 & 0.25 $\pm$ 0.04 & 0.13 $\pm$ 0.01 \\
        564.34 & 0.30 $\pm$ 0.05 & 0.18 $\pm$ 0.01 \\
        761.97 & 1.48 $\pm$ 0.23 & 0.90 $\pm$ 0.08 \\
        954.05 & 0.51 $\pm$ 0.08 & 0.31 $\pm$ 0.02 \\
        1101.18 & 0.64 $\pm$ 0.11 & 0.39 $\pm$ 0.04 \\
        1152.98 & 2.7 & 1.65 $\pm$ 0.15 \\
        1610.47 & 1.31 $\pm$ 0.20 & 0.80 $\pm$ 0.07 \\
        1914.70 & 2.0 $\pm$ 0.03 & 1.21 $\pm$ 0.11 \\
        \hline
    \end{tabular}     
\end{table}}

The total errors of the photoactivation yields (indicated 
as the point vertical bars in Fig.~\ref{FIG:8}) were defined 
as root-mean-square errors including counting 
statistics errors in $\gamma$-ray peak areas (in the 2\%
region for different $\gamma$-ray peaks), uncertainties of
detector efficiency (5\%), 
sample-to-detector distance (5\%), and the target 
dimensions (5\%) accuracies and are within (10-12)\%.
The uncertainties of the $\gamma$-ray intensities (indicated 
in the right column of Table~\ref{tbl2}) are smaller since they 
include only detector efficiency and peak area errors.
The intensities of 9 transitions, excluding 537.2 $keV$ one, 
turned out to be lower than those of NuDat 2.8 base
\cite{nudat2} by the average factor of 1.64 (0.10). 
537.2 $keV$ $\gamma$-ray intensity recalculation taking into
account different contributions of the ($\gamma$,n) and
($\gamma$,p) 
reactions gives the result lower by the factor of 1.92 (0.16).

In addition the solid and dashed curves in 
Fig.~\ref{FIG:8} represent 
the integral bremsstrahlung yields of the
$^{112}Sn(\gamma$,n)$^{111}Sn$ 
reaction calculated from the cross sections predicted by
the statistical 
theory of nuclear reactions implemented in NON-SMOKER
computer code \cite{Rauscher_2000} and TENDL data library 
\cite{tendl2019} respectively. Further interpretation of the 
$^{112}Sn(\gamma$,n)$^{111}Sn$ and
$^{112}Sn(\gamma$,p)$^{111m,g}In$ 
activation yields is currently underway.

\section{Conclusions}

Advantageous features of the 
$^{111}Sn \to ^{111}In \to ^{111}Cd$
radionuclide chain make it possible to determine intensities
of $\gamma$-ray transitions between the levels populated 
in the daughter $^{111}Cd$ nucleus by measuring and analyzing 
the yields of nuclear reactions 
leading to the formation of $^{111}Sn$ and 
$^{111}In$ nuclei. 
We applied this method to derive the updated values of 
10 strongest $\gamma$-ray transitions following 
$^{111}Sn$ radioactive decay and to derive
the values of integral yields of the 
$^{112}Sn(\gamma$,n)$^{111}Sn$ and
$^{112}Sn(\gamma$,p)$^{111}In$ photonuclear 
reactions in the near and above threshold energy range 
which are of interest 
for modeling the $\gamma$-scenario of the 
stellar nucleosynthesis.

New intensity values of $\gamma$-ray emissions following 
the $^{111}Sn$ 
nucleus decay will be of interest both for nuclear
spectroscopy theories and correct calculations of activation
cross sections and yields of those nuclear reactions
where the $^{111}Sn$ radioactive nuclide is a residual one.
The numerous relevant 
data presented in the EXFOR database \cite{exfor} 
have to be revised.    

\bibliography{bib}% Produces the bibliography via BibTeX.

\end{document}